\documentclass[12pt]{iopart}
\begin{document}
\title[Knot points of double--covariant system of
elliptic  equations]{Knot points of double--covariant system of
elliptic  equations and preferred frames in general relativity}
\author{ V Pelykh}
\address{\ Pidstryhach Institute Applied Problems in Mechanics and
Mathematics Ukrainian National Academy of Sciences, 3B Naukova
Str.,\\ Lviv, 79601, Ukraine}

\ead{pelykh@lms.lviv.ua}

\begin{abstract}
The elliptic system of equations, which is general-covariant and
locally $SU(2)$-covariant, is investigated. The new condition of
the Dirichlet problem solvability and the condition of zeros
absence for solutions are obtained for this system, which
contains in particular case the Sen-Witten equation. On this
basis it is proved the existence of the wide class of
hypersurfaces, in all points of which there exists a
correspondence between the Sen-Witten spinor field and
three-frame, which generalizes the Nester orthoframe. The Nester
special orthoframe also exists  on a certain subclass  containing
not only the maximal hypersurfaces.
\end{abstract}

\submitto{\JPA}

\maketitle
\section {Introduction}

The necessity for investigation of submanifolds, on which the
solutions of elliptic equations are equal to zero, is connected
with a fact that the necessary and sufficient conditions for
absence of such closed submanifolds of codimension one are
simultaneously the necessary and sufficient conditions for
uniqueness of the Dirichlet problem for these equations in the
domain.
Since the elliptic equations refer to the static solutions of the
given hyperbolic field equations, the non-uniqueness of solution
for the boundary value problem defines the non-stability of "zero
modes" of given field equations. Additionally
 there appears the necessity to study not only the closed
submanifolds and not only of codimension one , but all another ones, on which zeros of
solutions are located. This is related, in particular, to Sen-Witten equations (SWE),
for which the question about zeros existence had been discussed during a long time
$[1-4]$. The zeros absence for SWE solutions had been proved for the case when initial
data set for Einstein equations on the maximal hypersurface is asymptotically flat,
and  the local mass condition is fulfilled \cite{peljmp}. Since on maximal
hypersurface the system of equations splits into separate equations, the choice of
Cauchy surface as maximal enabled us to use the known results of investigations of
zeros for single equations  and to ascertain that in each point of maximal
hypersurface there exists two-to-one correspondence between Sen-Witten spinor and
Nester special orthonormal frame (SOF).

The purpose of this paper is to develop a new approach for
establishing the conditions of solvability and zeros absence for
general from the physical point of view elliptic systems of
equations. This will give the possibility to prove the existence
of the wide class of hypersurfaces, in all points of which there
exists the two-to-one correspondence between Sen-Witten spinor
and a certain three-frame; we will name it  Sen-Witten
orthonormal frame (SWOF). In all points on such hypersurfaces
there exist also the well defined lapses and shifts, associated
by Ashteckar and Horowitz \cite{ash1} with Sen-Witten spinor. On
a subclass of this class, including also the maximal
hypersurfaces, we establish the existence of two-to-one
correspondence between Sen-Witten spinor and Nester three-frame.

\section{Preliminaries}

We introduce first three definitions.

{\it Definition 1. The knot point of the component of the
solution is a point, in which the
 component is equal zero.}

 {\it Definition 2. The knot point of the solution for the elliptical
 system of equations is  a
point, in which the solution is equal to zero.}

 From the general theory of elliptic differential equations is known that nontrivial
solutions cannot vanish on an open subdomain, but they can turn to zero on subsets of
lower dimensions $k, k=0,1,...n-1$, where $n$ is dimension of the domain.

{\it Definition 3. The knot submanifold of dimension $s, s=1,2,...n-1,$
 is a maximal connected subset\footnote[1]{Maximal connected subset
 $A$ is a nonempty connected subset such that the only connected
subset containing $A$ is $A$.}of dimension $s$
  consisting of knot points of the solution.}

   Discrete set of knot
points is $0$-submanifold. We will show in Sec.$\,3$ that for the
system of differential equations, interesting for us, all knot
subsets are formed by intersection of knot surfaces of the
components of the solution.

The connection between the unique solvability for the boundary
value problem in ${\bf R}^n$ and absence of $(n-1)$-dimensional
 closed knot submanifolds
was established by Picone \cite{pic1, pic2}. The existence of such connection follows
from the next
 consideration: if the boundary value problem in a certain domain
$\Omega$ is uniquely solvable, then the boundary value problem is
also uniquely solvable for any subdomain $\Omega_1\subseteq
\Omega$. This excludes the possibility of existence of nontrivial
solutions which turn to zero on the boundary of the arbitrary
domain $\Omega_1$, i.e., excludes the possibility of existence of
 closed knot submanifolds of codimension one, and vice versa.

 The known investigation of elliptical equations of general form does
 not allow to obtain the conditions for all knot points absence.
 For example, even in the case of the only single equation of
 general form it is proved the absence of zeros  only of infinite order
\cite{aro, cor}. That is why further we will examine only such
 general equations, which possess also the necessary physical
 properties, in particular, symmetry properties.

Let $\Omega$ be a bounded closed spherical-type domain in
three-dimensional Riemannian space $V^3$,
otherwise, (i) its boundary $\partial \Omega$ in every point have a
tangent plane; (ii) for every point $P$ on the boundary there
exists a sphere, which belongs to $\Omega$, and the boundary
of sphere includes  the point $P$.

In the domain $\Omega$ let us consider the system of elliptic second order
equations

\begin{equation}\label{1}
\frac1{\sqrt{-h}}\frac{\partial}{\partial
x^\alpha}\left(\sqrt{-h}h^{\alpha\beta}\frac{\partial}{\partial
x^\beta}u_A\right)+ C_A{}^Bu_B=0,
\end{equation}
where $h^{\alpha\beta}$ ---  components of the metric tensor in
$V^3$; they are   arbitrary real functions of independent real
variables $x^\alpha$, continuous in $\Omega$ and
 the quadratic form
$h^{\alpha\beta}\xi_\alpha\xi_\beta$ is  negative definite. The unknown functions
$u_A$
of independent variables $x^\alpha$
 are  the elements of complex vector space ${\bf C}^2$, in which
the skew symmetric tensor $\varepsilon^{AB}$ is defined, and the
group $SU(2)$ acts. $C_A{}^B$ is Hermitian $(1,1)$
spinorial tensor.

The system of equation (\ref{1}) is covariant under the arbitrary
transformations of coordinates in $V^3$, and covariant under the
local $SU(2)$--transformations of unknown functions in a local
space isomorphic to the complexified tangent space in every point
to $V^3$.

 Picone had ascertained that at
arbitrary coefficients of elliptic equations the boundary value problem is uniquely
solvable, and the closed knot submanifolds  of codimension one are absent,
respectively, only in the domains with enough small intrinsic diameter.

The general conditions for the absence of closed knot surfaces for strong elliptic
system (\ref{1}) are ascertained by Theorem 1 \cite{sko1}.

{\bf Theorem 1.} {\it If in domain $\Omega$ there exist
symmetrical quadratic functional second-order matrices $B_1, B_2,
B_3$ of $C^1$ class,  such that matrix

\[ \sqrt{-h}C-\sum_{\alpha=1}^3\frac{\partial B _\alpha}{\partial
x^\alpha}+ B^TG^{-1}B \]
is  positive definite, where $B=(B_1, B_2, B_3),
G=\sqrt{-h}\, {\rm diag}(\|h_{\alpha\beta}\|, \|h_{\alpha\beta}\|,
\|h_{\alpha\beta}\|$, then the solutions of system of equations (\ref{1}) with matrix
$C=\|C_A{}^B\|$ of $C^1$ class
 do not have the closed knot surfaces in domain
$\Omega$}.

 The effective geometrical conditions of
B-matrix existence and corresponding unique solvability  of Dirichlet problem in
dependence on the domain intrinsic diameter were obtained \cite{sko1} for the
Euclidean  space. Such conditions are important, for example, in the theory of nuclear
reactors. Since the conditions of knot manifolds absence for quantum fields equations
are the point of our interest,  we  further will concentrate our attention on the
conditions of  knot points absence in the domains of arbitrary as well as infinite
intrinsic diameter.

Evidently, if matrix $C$ is  positive definite, then the conditions of Theorem 1 are
fulfilled for $B\equiv0$, and closed knot surfaces are absent in the domain with
arbitrary intrinsic  diameter. Simultaneously, the boundary value problem for the
system of equations (\ref{1}) is uniquely solvable.

Theorem 1 does not indicate the conditions at which  knot points, lines as well as all
knot surfaces for solutions of equations (\ref{1}) are absent. We will obtain them in
 Sec.$\,3$.

\section{Conditions for the absence of knot points}

 In the case of a single selfadjoint elliptic equation in $V^3$
 the knot submanifolds can be only the surfaces
which divide the domain, but in the case of a system of equations the topology of knot
submanifolds becomes more various: it can be also the lines and the points.
 We can take this fact into account and ascertain the
conditions for the knot manifolds absence exploiting  the double covariance of the
system of equations (\ref{1}) and using Zaremba-Giraud Lemma, generalized at first by
Keldysh and Lavrentiev \cite{kel} and later by Oleynik \cite{ole}.

Let us introduce the matrix \[ R:=||R_{A^\prime}{}^B||:=
\left(\begin{array}{cc} \alpha & \beta
\\ -\overline{\beta} & \overline{\alpha}
\end{array}\right),
\qquad \alpha\overline\alpha+\beta\overline\beta=1,
\]
which is of the group  $SU(2)$, and let its elements additionally
 satisfy the condition
\[C_0{}^1\beta^2+(C_0{}^0-C_1{}^1)\alpha\beta-C_0{}^1\alpha^2=0.
\] Therefore,
\[C_{0^\prime}{}^{1^\prime}=R_{0^\prime}{}^AC_A{}^BR^{1^\prime}{}_B=0,
\] and in accordance with  Hermicity of  matrix $C$ also
$\overline C_{0^\prime}{}^{1^\prime}=C_{1^\prime}{}^{0^\prime}$.
Then $C_0:=C_{0^\prime}{}^{0^\prime}$ and
$C_1:=C_{1^\prime}{}^{1^\prime}$ are eigenvalues of  matrix
$C=||C_A{}^B||$. This follows from a fact that for arbitrary
matrix $R\in SU(2)$ the identity
 \[-\varepsilon R \varepsilon\equiv R^{T+}\]
 is valid, where $\varepsilon=\|\varepsilon^{AB}\|$. Therefore \[C^\prime=-\varepsilon R
\varepsilon C R^{T+}=R^{T+} C R^T={\rm diag}(C_{0},\; C_{1}).\]

Let us denote
\[\Delta:=C_1{}^1-C_0{}^0-\left[\left(C_1{}^1-C_0{}^0\right)^2+
4|C_0{}^1|^2\right]^{1/2},\] and let us denote by $S$ a set of
points in domain $\Omega$, in all points of which $C_0{}^1$ does
not equal to zero, and let us denote by $T$ a set of points, in
which $C_0{}^1$ is equal to zero. Then the elements of the matrix
 $R$, which transforms the matrix $C$ to diagonal form , satisfy
  on the set $S$ the conditions
\[\alpha\overline\alpha(1+\Delta^2/4|C_0{}^1|^2)=1,\qquad
\beta=\alpha\Delta/2C_0{}^1\] and on the set $T$ the conditions
\[\alpha\overline\alpha=1,\qquad\beta=0.\]
 Functions $u_{0^\prime}$
and $u_{1^\prime}$ on the set $S$ will be following:
\begin{equation}
 u_{0^\prime}=\overline\alpha\left(u_0+\frac{\Delta}
{2{\overline C}_0{}^1}\,u_1\right),\qquad
u_{1^\prime}=\alpha\left(-\frac{\Delta }{2C_0{}^1}\,u_0
+u_1\right),\label{2}
\end{equation}
 and on set $T$ they will be
\begin{equation}
u_{0^\prime}= \overline\alpha u_0, \qquad
 u_{1^\prime}=\alpha u_1.\label{3}
\end{equation}
 Respectively, eigenvalue $C_0$ on $S$ is:
\[C_{0}={ 4C_0{}^0|C_0{}^1|^4+\left(4
\Delta\,|C_0{}^1|^2+C_1{}^1\Delta^2\right)
\left(4|C_0{}^1|^2+\Delta^2 \right)\over
4|C_0{}^1|^2\left(4|C_0{}^1|^2+\Delta^2 \right)}\]
 and coincides
with $C_0{}^0$ on the set $T$.

{\bf Lemma.} {\it If real and imaginary parts of functions $u_A$ and of elements of
matrix $C_A{}^B$ are functions of class $C^2$ in domain $\Omega$, then the real and
imaginary parts of functions $u_{A^\prime}$ defined by conditions (\ref{2})-(\ref{3})
are also the functions of class $C^2$ in this domain.}

{\it Proof.} Taking into account that it is always  possible to
choose ${\rm Im}\,\alpha\!\in C^2(\Omega)$, from direct
calculation we obtain that on a set $S$ there are exists first
and second derivatives of real and imaginary parts of functions
$u_{A^\prime}$ and $\alpha$ with respect to arguments
$(\Delta^2/4|C_0{}^1|^2)$ and $(\Delta/2|C_0{}^1|^2)$ and that

\fl\[\lim_{P\ni S\rightarrow Q\in T}{\rm Re}\,\alpha
^{(m)}(P)={\rm Re}\,\alpha ^{(m)}(Q),\hspace{5mm} \lim_{P\ni
S\rightarrow Q\in T}{\rm Im}\,\alpha ^{(m)}(P)={\rm Im}\,\alpha
^{(m)}(Q),\] \[ \lim_{P\ni S \rightarrow Q\in T}{\rm Re}\,
u_{A^\prime}^{(m)}(P)={\rm Re}\, u_{A^\prime}^{(m)}
(Q),\hspace{5mm} \lim_{P\ni S\rightarrow Q\in T}{\rm Im}\,
u_{A^\prime}^{(m)}(P)={\rm Im}\, u_{A^\prime}^{(m)}(Q),\]
 where
 symbol $f^{(m)}$ denotes arbitrary  partial derivatives of
 order $m=0,1,2$.

The following theorem is valid.

{\bf Theorem 2.} {\it Let:

a) real and imaginary parts of the elements of  matrix $C$ be of
$C^2$ class in the domain $\Omega$;

 b) at
least one eigenvalue of matrix $C$, for definiteness $C_0$, is
non-negative everywhere in $\,\Omega$;

c) real or imaginary part of the function
 \[v:= \begin{cases}
     {\left(u_0+\frac{\Delta} {2{\overline
C}_0{}^1}\,u_1\right)\left|_{S\bigcap\partial\Omega}\right., &{}
\\
   u_0\left|_{T\bigcap\partial\Omega}\right. &{}}
  \end{cases}
\]
 does not equal to zero in any point.

Then solution $u_A$ of class $C^2$ for the system of equations
(\ref{1}) does not have any knot points in
 the domain $\Omega$ of spherical type.}

{\it Proof:} The system of equations $(\ref{1})$ is covariant under the arbitrary
transformations of coordinates and under the local transformations from the group
$SU(2)$ that allows to use them independently. Let us apply on the first step the
$SU(2)$ spinor transformation $u_{A}\rightarrow R_{A^\prime}{}^Bu_B$, which transforms
the matrix $C$ to the diagonal form, and under which the equation (\ref{1}) is
covariant.

The eigenvalues of matrix $C$ are real, therefore, the resulting system of
equations (\ref{1})  splits into a system
of four independent equations for real and imaginary parts of spinor
$u_{A^\prime}$. Taking into account that $u_{A^\prime}$, $C_{0}$ and
 $C_{1}$ are scalars under  transformations of
coordinates, and $C_0{}\geq0$, we can apply the Zaremba-Giraud
principle in the general form grounded by Oleynik \cite{ole} to
every  equation containing $C_{0}$. According to this principle,
if in a certain point $P_0$ on the sphere the nonconstant
function in the ball turns to zero, and everywhere in the ball
${\rm Re} \,u_{0^\prime}<0$, then $\left<{\rm d\,
Re}\,u_{0^\prime},\,\,l\right>_{|_{P_0}}<0$. Here $l$ --
 arbitrary vector field, for which
$\left<n,\,\,l\right>_{|_{P_0}}>0$, and $n$ is one-form of
intrinsic normal to the sphere in the point $P_0$.

Let us show further that a set of the knot points for function
${\rm Re}\,u_{0^\prime}$ does not contain the isolated points. Let us
assume that such point exists, i.e. ${\rm Re}\,u_{0^\prime}=0$, and in
a certain neighborhood of the point $P_0$ the function has a constant
sign. For definiteness let in this neighborhood be
$u_{0^\prime}<0$.
 Let us
consider a sphere, on which the point $P$ lies and is so small
that completely belongs to the mentioned neighborhood of the
point $P$. Then, using Zaremba-Giraud principle, we obtain
$\left<{\rm d\,Re}\,u_{0\prime},\,\,n\right>{|_{P_0}}>0$, and
therefore in any neighborhood of the point $P_0$, located outside
the ball the function ${\rm Re}\,u_{0^\prime}$ changes its sign,
and that is why its zeros are not isolated. Therefore, they form
the surfaces which divide $\Omega$. Since  $C_0\geq0$ , then it
follows from the maximum principle that the closed knot surfaces
for the components of solution ${\rm Re}\,u_{0^\prime}$ are
absent. Analogous conclusion is true also for the component of
solution ${\rm Im}\,u_{0^\prime}$. This means that the only
surfaces having common points with the boundary of domain
$\Omega$ can be the knot surfaces of real or imaginary part of
function $u_{0^\prime}$. According to condition c), if, for
definiteness,

\[{\rm Re} \left(u_0+\frac{\Delta} {2{\overline
C}_0{}^1}\,u_1\right)| _{S\bigcap\partial\Omega}\neq0,\qquad
 {\rm Re}\,u_0|_{{T\bigcap\partial\Omega}}\neq0,
\] then we can choose \[{\rm
Re}\,\overline\alpha|_{S\bigcap\partial\Omega}\neq\!\left\{\left[{\rm
Re}\, \left(u_0+\frac{\Delta} {2{\overline
C}_0{}^1}\,u_1\right)\right]^{-1}{\rm
Im}\,\overline\alpha\,\,{\rm Im}\, \left(u_0+\frac{\Delta}
{2{\overline
C}_0{}^1}\,u_1\right)\!\right\}\!|_{S\bigcap\partial\Omega},\]
\[{\rm
Re\,}\overline\alpha|{_{T\bigcap\partial\Omega}}\neq\left[\left({\rm
Re}\, u_0\right)^{-1}{\rm Im}\,\overline\alpha\,{\rm Im}\,
u_0\right]|_{T\bigcap\partial\Omega}\]
 and obtain
  \[ \left[{\rm
Re}\,\overline\alpha\,{\rm Re} \left(u_0+\frac{\Delta}
{2{\overline C}_0{}^1}\,u_1\right)-{\rm
Im}\,\,\overline\alpha\,\,{\rm Im} \left(u_0+\frac{\Delta}
{2{\overline
C}_0{}^1}\,u_1\right)\right]|_{S\bigcap\partial\Omega}\]\[\equiv
{\rm Re}\, u_{0^\prime}|_{S\bigcap\partial\Omega}\neq0,\]
 \[
\left({\rm Re\,}\overline\alpha\,\,{\rm Re}u_0 -{\rm
Im}\,\overline\alpha\,\,{\rm Im}\,
u_0\right)|_{T\bigcap\partial\Omega}\equiv {\rm Re}\,
u_{0^\prime}|_{T\bigcap\partial\Omega}\neq0. \]

 Therefore  knot surfaces
as well as lines and points
 of the real (or imaginary) part are absent, and that is why any
 knot points of complete solution $u_A$ are also absent. The
 statement of the theorem is proved.

Note. If the conditions a) and b) of the Theorem be fulfilled, and the matrix $C$ be
non-negative definite in domain $\Omega$, then both eigenvalues are non-negative, and,
therefore, the
 boundary value problem for the system of equations
(\ref{1})  is uniquely solvable in arbitrary bounded domain, as it
follows from the classical maximum principle. Otherwise
 the solution in finite domain exists
only in the case when its intrinsic diameter does not overcome a
certain value.

\section{The conditions of knot points absence for the solutions of
Sen-Witten equation }

 After Witten's positive energy proof the attempts of development of tensor method
for proof were performed along two lines. The attempts of the tensor interpretation
for Sen-Witten spinor field belong to the first line. In particular, Ashtekar and
Horowitz [1] used Sen-Witten spinor field for determination of a class of preferred
lapses $T:=\lambda$ and shifts $ T^a:=-\sqrt2\,i\lambda^{+(A}\lambda^{B)}$. Dimakis
and M${\rm {\ddot u}}$ller-Hoissen \cite{dim1, dim2} had defined a preferred class of
orthonormal frame fields in which spinor field take a certain standard form.
Frauendiener \cite{fra} had noticed                 a correspondence between
Sen-Witten spinor field and a triad. But, as it was shown by Dimakis and M${\rm {\ddot
u}}$ller-Hoissen, frame fields cannot exist in the knot points of the spinor field.

Among the works of the second line the most developed is Nester's method which is
grounded on the new gauge conditions for the special orthonormal frame (SOF):
\begin{equation}\label{4}
*q:=\varepsilon^{abc}\omega_{abc}=0,\qquad\widetilde q
_b:=\omega^a{}_{ba}=F_b,
\end{equation}
where $\omega_{abc}$ are the connection one forms coefficients
and $F$ is arbitrary everywhere on $\Sigma_t$ defined exact
one-form.

 An essential part of Witten's proof of nonnegativity for ADM mass is application of
Sen-Witten equation (SWE)
\begin{equation}\label{5}
{\cal D}^B{}_C\beta^C=0
\end{equation}
 with appropriate asymptotic conditions on the spacelike hypersurface $\Sigma$
in four-dimensional Riemannian manifold $M=\Sigma\times R $ with each
$\Sigma_t=\Sigma\times \{t\}$ spacelike. Initial data set $(\Sigma_t, h_{\mu\nu},
{\cal K}_{\pi\rho})$ satisfies the constraints, and is asymptotically flat.
 An action
of  operator ${\cal D}_{AB}$ on spinor fields is
\[ {\cal D}_{AB}\lambda_C =D_{AB}\lambda_C+\frac{\sqrt2}{2}{\cal
K}_{ABC}{}^D\lambda_D, \] where $D_{AB}$ --- spinorial form of
the derivative operator $D_\alpha$ compatible with the metric
$h_{\mu\nu}$ on $C^\infty$ hypersurface $\Sigma_t,\;\; {\cal
K}_{ABCD}$ --- spinorial
 tensor of extrinsic curvature of hypersurface $\Sigma_t$.

The existence and uniqueness theorem for solution of equation (\ref{5}) in
corresponding Hilbert space with some asymptotic conditions was proved by Reula
\cite{reu} (see also \cite{ash1}).

 Let us
ascertain the conditions of zeros absence for these solutions on $\Sigma_t$ using the
results of Sec.$\,2$. From equation (\ref{5}), taking into account the equation of
Hamiltonian constraint on $\Sigma_t$,
 in Gauss normal coordinates we obtain \cite{peljmp}:
 \[
 {\cal D}_A{}^B{\cal
D}_{BC}\lambda^C=\frac1{2\sqrt{-h}}
 \frac{\partial}{\partial
 x^\alpha}\left(\sqrt{-h}h^{\alpha\beta}\frac{\partial}{\partial
 x^\beta}\lambda_A\right)-\frac{\sqrt2}{2}{\cal K}D_{AB}\lambda^B-
\]
\begin{equation}\label{6}
-\frac{\sqrt2}{4}\lambda^BD_{AB}{\cal K}+\frac14{\cal K}^2\lambda
_A+\frac18{\cal K}_{\alpha\beta}{\cal K}^{\alpha\beta}\lambda_A
+\frac14\mu\lambda_A=0.
\end{equation}
Therefore, the system of equations (\ref{6}) is a system of the form (\ref{1}); if it
does not have the knot points, the SWE also does not have them.

Spinorial tensor
\begin{equation}\label{7}
C_A{}^B:=\frac{\sqrt2}{4}D_{A}{}^{B}{\cal K}+\frac14 \varepsilon
_A{}^B\left(2{\cal K}^2+\frac12{\cal K}_{\pi\rho}{\cal
K}^{\pi\rho}+\mu\right)
\end{equation}
is Hermitian because $({\cal D}_A{}^B{ \cal
K})^+=\left(\varepsilon^{BC}{\cal D}_{AC}{\cal K}\right)^+=
\left(\varepsilon^{BC}\right)^+\left({\cal D}_{AC}{\cal
K}\right)^+=-\left({\cal D}_{AC}{\cal
K}\right)\varepsilon^{CB}\\=({\cal D}_A{}^B{\cal K})$.

So, the SWE solutions of class $C^2$ do not
have the knot points  in a bounded closed domain $\Omega$
of spherical type on $\Sigma_t$, if for spinorial tensor $C_A{}^B$ in this
domain and for the boundary values of the solution the conditions
 of Theorem 2 are
fulfilled.

Let us consider further a sequence $\Omega_n$ of increasing
domains of spherical type covering $\Sigma_t$. If in every domain
the conditions of Theorem 2 are fulfilled, then all solutions of
class $C^2$ do not have the knot points in $\Omega_n$. According
to Reula, on $\Sigma_t$ there exists the SWE solution of
$\lambda^C=\lambda^C_{\infty}+\beta^C$ form, where
$\lambda^C_{\infty}$ is asymptotically covariant constant spinor
field on $\Sigma_t$, $\beta^C$ is an element of  Hilbert space
 ${\cal H}$, which is the Cauchy
completion of $C_0^\infty$ spinor fields under the norm
\[\mid\mid\beta^E\mid\mid^2_{\cal
H}=\int\limits_{\Sigma_t}\left({\cal
D}^A{}_B\beta^B\right)^+\left({\cal D}_{AC}\beta^C\right)dV.\]

 Solution $\lambda^C$ belongs properly to $C^\infty$
class. From the asymptotical flatness condition it follows that
$(\Delta^2/(4|C_0{}^1|^2)$, as well as real and imaginary parts
of functions $(\Delta/2{\overline C}_0{}^1)$ and
 $(\Delta/2 C_0{}^1)$ vanish asymptotically. Therefore,
 condition c)  from Theorem 2 asymptotically take a form:\;
 ${\rm Re}\, \lambda^0_\infty\neq0$ or ${\rm Im}\, \lambda^0_\infty\neq0$.
 In such a way we obtain the following theorem:

{\bf Theorem 3.} {\it Let:

a) initial data set be asymptotically flat;

b) everywhere on $\Sigma_t$ the matrix of spinorial tensor (\ref{7}) have at least one
non-negative eigenvalue, for definiteness $C_0$;

c) ${\rm Re}\,\lambda^0_{\infty}$ or ${\rm
Im}\,\lambda^0_{\infty}$ asymptotically nowhere equal zero.

Then the asymptotically constant nontrivial solution $\lambda^C$ to SWE does
not have the knot points on $\Sigma_t$.}

The conditions of Theorem 3 are fully admissible  from the
physical point of view.

\section{Towards Sen-Witten equation, special
orthonormal frame and preferred time  variables}

Usually the question about existence of system of coordinates or
orthonormal basis, which satisfy certain  gauge conditions, is
reduced to the question about existence of solution for nonlinear
system of differential equations and often can be solved only at
some additional limitations and assumptions \cite{nes1}.

The existence theorem for Sen-Witten (linear) equation and the
 Theorem 3 about their
zeros (Sec.$\,4$) on surfaces, which can be not  maximal, allow
us to prove  the existence   of a certain class of orthonormal
three-frames in all points of these hypersurfaces which satisfies
gauge conditions.
\[
\varepsilon^{abc}\omega_{abc}\equiv*q=0,\qquad\omega^a{}_{1a}\equiv-\widetilde
q _1=F_1,\qquad \omega^a{}_{2a}=-\widetilde q_2=F_2,\]
\begin{equation}\label{8} \omega^a{}_{3a}
=-\widetilde q_3={\cal K}+F_3,
\end{equation}
 and generalizes Nester's SOF. Such three-frame we will name as Sen-Witten
orthonormal frame (SWOF)

{\it Theorem 4.  Let the conditions of Theorem 3 be fulfilled. Then everywhere on
$\Sigma_t$ there exists a two-to-one correspondence between Sen-Witten spinor and
Sen-Witten orthonormal frame.

Proof}:

Really, let all conditions of Theorem 3 be fulfilled on
$\Sigma_t$. Then SWE solution $\lambda_A$ does not have the knot points
anywhere on $\Sigma_t$.
 This allows to prove
on such $\Sigma_t$ the Sommers \cite{som} assumption
 that spatial
null one-form $L=-\lambda_A\lambda_B$ on $\Sigma_t$ is nonzero,
and allows to turn everywhere on $\Sigma_t$ to the "squared" SWE
represented in the form:
\begin{equation}\label{9}
\left<{\widetilde L},\,D\otimes L\right>-{\cal
K}L+3!\,i*\left(n\land D\land L\right)=0,
\end{equation}
where $\:\left<\widetilde L,\,D\otimes L\right>$ is one--form with
components $\widetilde L_\nu D_\mu L^\nu,\: \widetilde L=\mid
L\mid^{-1}*\left(L\land\overline L\right)$ is nonzero spatial
one--form, and $n$ is one form of unit normal to $\Sigma_t$.

The bilinear form \[\frac1{\sqrt2}n^{A{\dot A}}\lambda_A\overline{\lambda}_{\dot
A}=\lambda_A\lambda^{A+}\equiv \lambda,\] where $n$ is one-form of a unit normal to
$\Sigma_t$, is Hermitian positive definite one, and the solution
 $\lambda_A$ does not have the knot points on $\Sigma_t$.
Consequently, we can further introduce real nowhere degenerated
orthonormal 4-coframe $\theta^m$ as
\begin{equation}\label{10}
\theta^0\equiv n=N{\rm d}t,\quad
\theta^1=\frac{\sqrt2}{2\lambda}(L+\overline L),\quad
\theta^2=\frac{\sqrt2}{2\lambda i}(L-\overline L),
\quad\theta^3=\widetilde L
 \end{equation} and represent
immediately (\ref{9}) in the form
\begin{equation}\label{11}
-\left<\theta^1,\,D\otimes \theta^3\right>-{\cal
K}\theta^1+3!\,*\left[n\land(D+F)\land\wedge \theta^2\right]=0,
\end{equation}
\begin{equation}\label{12}
\left<\theta^2,\,D\otimes \theta^3\right>+{\cal K}\theta^3+3!\,*\left[n\land(D+F)\land
\theta^1\right]=0,
\end{equation}
where $F=D\ln \lambda$. The system of equations
(\ref{11})-(\ref{12}) includes only four independent equations,
and they are equations (\ref{8}) for the connection one-forms
coefficients. From this it follows that if on $\Sigma_t$ the
conditions of Theorem 3 and SWE are fulfilled, then on $\Sigma_t$
there exists  three-frame $\theta^a$ defined by (\ref{10}) in
which conditions (\ref{8}) are fulfilled.

Inversely, if on $\Sigma_t$ in some three-frame $\theta^a$ the conditions of
  Theorem 3
and conditions (\ref{8}) are fulfilled, then it follows from
condition of Theorem 3 that these one-forms have a form
$\theta^a=\theta^a_\infty+\phi^a$, where $\theta^a_\infty$ tend
asymptotically to the covariant constant forms and $\phi^a$
belongs to ${\cal H}$.
 We can turn from four-frame
$\theta^m\equiv\{n,\, \theta^a\}$ to one-forms
$\theta^0,\,L,\,\widetilde L$, assuming
 $\lambda_A
|_ {{\scriptstyle{\Sigma}_{\scriptscriptstyle t}}}\neq0$.
  After this we obtain equation (\ref{9})
and further (\ref{5})\footnote[1]{The equivalence of the SWE (\ref{5}) and of the
equation (\ref{6}) is proved by Reula \cite{reu}.} for spinor field $\lambda^A$,
which, as we have demonstrated previously, indeed does not have the knot points on
selected hypersurface $\Sigma_t$ and which together with asymptotical conditions
defines up to sign the spinor field $\lambda^A$.  Mentioned in conditions of the
Theorem correspondence between  Sen-Witten spinor field and Nester's SOF is defined by
relationship (\ref{10}).

We have proved \cite{peljmp} that if initial set $(\Sigma_t, h_{\mu\nu}, {\cal
K}_{\pi\rho})$ on maximal hypersurface\footnote[2]{Maximal surfaces are spacelike
submanifolds af a Lorentzian manifold which locally maximize the induced area
functional.} $\Sigma_t$ is asymptotically flat and satisfies the dominant energy
condition, then everywhere on $\Sigma_t$  from existence Sen-Witten spinor field
follows existence Nester's three-frame and conversely.
  Theorem 3 allows to strengthen
significantly this result taking away  the assumption that
 $\Sigma_t$ is maximal. Indeed, if all conditions of Theorem 3
are fulfilled on $\Sigma_t$, and additionally  the one-form ${\cal K}\widetilde L$ is
globally exact, we can perform in conditions  the identification $F \equiv {\rm
d}\ln\lambda+{\cal K}\theta^3$ and obtain the Nester's gauge (\ref{4}), or to perform
the inverse transition --- from Nester's gauge to SWE. Therefore, if on $\Sigma_t$ the
conditions of Theorem 3 are fulfilled, then SWE and Nester's gauge are equivalent if
and only if the one-form ${\cal K}\lambda^{+(A}\lambda^{B)}$ is exact.  In this case
the correspondence between Sen-Witten spinor and Nester's SOF is also ascertained by
relationship (\ref{10}).

 Ashteckar
and Horowitz \cite{ash1} have accented on the necessity of zeros investigations for
SWE solutions introducing the vector interpretation of Sen-Witten's spinor which
defines a preferred lapse and shift. Evidently, the fulfilling of the Theorem 3
conditions ensures the existence of corresponding lapses and shifts well defined
everywhere on $\Sigma_t$.  And also the preferred class of orthonormal four-frame
fields introduced by Dimakis and  M${\mathrm {\ddot u}}$ller-Hoissen exists in all
points of $\Sigma_t$ under fulfilling of the Theorem 3 conditions.

\section{Conclusions}

The presence of zeros in the solutions of elliptic equations is
rather ordinary than exceptional case, therefore, it is necessary
to prove the absence of zeros for concrete cases.

The represented investigation demonstrates the possibility for
obtaining the condition of the knot manifolds absence for enough
general system of elliptic second order equations owing to its
double covariance.

The application of this result to SWE allows to prove the equivalence of SWE and gauge
conditions (\ref{8}), and, respectively, the existence of an everywhere well defined
two-to-one correspondence  between  Sen-Witten spinor field and the SWOF, which is the
Nester SOF in the particular case, when  one of the one-forms ${\cal K}\theta^a$ is
exact. Therefore, the indicated correspondence exists not only on the unique
--- maximal
--- hypersurface, but on the whole set of asymptotically flat
hypersurfaces.

Ashteckar and Horowitz \cite{ash1} had shown that the Reula results hold even if the
energy condition is mildly violated. Also the conclusion about existence of  special
three- and four-frames, as well as preferred lapses and shifts, is stable under  the
violation of the energy condition, because, as it is seen from (\ref{7}), there
 exist the hypersurfaces, on which this condition of knot
points absence is fulfilled at violation of the energy condition.

 \end{document}